\pdfoutput=1

\documentclass[final,3p,11pt]{elsarticle}

\usepackage{graphicx}
\usepackage{amssymb}

\usepackage{lineno}

\usepackage{framed}

\immediate\write18{%
makeindex -s nomencl.ist -o \jobname.nls -t \jobname.nlg \jobname.nlo%
}

\usepackage{nomencl}
\makenomenclature
\usepackage{etoolbox}
\renewcommand\nomgroup[1]{%
  \item[\bfseries
  \ifstrequal{#1}{A}{\small{Symbols}}{%
  \ifstrequal{#1}{G}{\small{Greek symbols}}{%
  \ifstrequal{#1}{S}{\small{Subscripts}}{}}}%
]}

\usepackage{multicol} 

\makenomenclature

\renewcommand*\nompreamble{\begin{multicols}{2}}

\renewcommand*\nompostamble{\end{multicols}}

\journal{-}

\begin{document}

\begin{frontmatter}


\title{Deposition of particle pollution in turbulent forced-air cooling}


\author[label1]{Jason Stafford\corref{cor1}}
\address[label1]{School of Engineering, University of Birmingham, UK}
\cortext[cor1]{Corresponding author: j.stafford@bham.ac.uk}

\author[label2]{Chen Xu}
\address[label2]{Nokia, Murray Hill, NJ 07974, USA}



\begin{abstract}
Rotating fans are the prevalent forced cooling method for heat generating equipment and buildings. As the concentration of atmospheric pollutants has increased, the accumulation of microscale and nanoscale particles on surfaces due to advection-diffusion has led to adverse mechanical, chemical and electrical effects that increase cooling demands and reduce the reliability of electronic equipment. Here, we uncover the mechanisms leading to enhanced deposition of particle matter (PM$_{10}$ and PM$_{2.5}$) on surfaces due to turbulent axial fan flows operating at Reynolds numbers, $Re \sim 10^5$. Qualitative observations of long-term particle deposition from the field were combined with \textit{in situ} particle image velocimetry on a telecommunications base station, revealing the dominant role of impingement velocity and angle. Near-wall momentum transport for $10 < y^+ < 50$ were explored using a quadrant analysis to uncover the contributions of turbulent events that promote particle deposition through turbulent diffusion and eddy impaction. By decomposing these events, the local transport behaviour of fine particles from the bulk flow to the surface has been categorised. The transition from deposition to clean surfaces was accompanied by a decrease in shear velocity, turbulent stresses, and particle sweep motions with lower flux in the wall-normal direction. Finally, using these insights, selective filtering of coarse particles was found to promote the conditions that enhance the deposition of fine particle matter.

\end{abstract}


\end{frontmatter}

\begin{table*}[!t]
\begin{framed}
\nomenclature[A]{$u,v$}{velocity components [m s$^{-1}$]}
\nomenclature[A]{$U$}{velocity [m s$^{-1}$]}
\nomenclature[G]{$\omega$}{rotational speed [rad s$^{-1}$]}
\nomenclature[A]{$Re$}{Reynolds number $\equiv\frac{\rho{u_{b} r_{b}}}{\mu}$ [-]}
\nomenclature[A]{$Pe$}{Peclet number $\equiv\frac{6\pi\mu{r_p{^2}}u}{k_B T}$}
\nomenclature[A]{$St$}{Stokes number $\equiv\frac{\rho_{p} d_{p}^{2} \sqrt{\varepsilon}}{18 v^{3 / 2} \rho}$ [-]}
\nomenclature[A]{$Sc$}{Schmidt number $\equiv\frac{\nu}{D}$ [-]}
\nomenclature[A]{$T$}{temperature [K]}
\nomenclature[A]{$k_B$}{Boltzmann constant [m${^2}$ kg s$^{-2}$ K$^{-1}$]}
\nomenclature[A]{$r$}{radius [m]}
\nomenclature[A]{$d$}{diameter [m]}
\nomenclature[A]{$D$}{diffusion coefficient [m$^2$ s$^{-1}$]}
\nomenclature[A]{$m$}{mass [kg]}
\nomenclature[A]{$J$}{flux density [kg m$^{-2}$ s$^{-1}$]}
\nomenclature[A]{$c$}{concentration [kg m$^{-3}$]}
\nomenclature[A]{$A$}{area [m$^2$]}
\nomenclature[A]{$t$}{time [s]}
\nomenclature[A]{$x,y,z$}{coordinates [m]}
\nomenclature[A]{$y^{+}$}{wall distance $\equiv\frac{u_{\tau} y}{\nu}$ [-]}
\nomenclature[A]{$u_{\tau}$}{shear velocity $\equiv\sqrt{\frac{\tau}{\rho}}$ [m s$^{-1}$]}
\nomenclature[A]{$Q$}{volumetric flow rate [m$^3$ s$^{-1}$]}
\nomenclature[A]{$p$}{pressure [Pa]}
\nomenclature[A]{$R$}{thermal resistance [K W$^{-1}$]}
\nomenclature[G]{$\rho$}{density [kg m$^{-3}$]}
\nomenclature[G]{$\mu$}{dynamic viscosity [kg m$^{-1}$ s$^{-1}$]}
\nomenclature[G]{$\nu$}{kinematic viscosity [m$^2$ s$^{-1}$]}
\nomenclature[G]{$\tau$}{wall shear stress [Pa]}
\nomenclature[G]{$\tau^+$}{particle relation time $\equiv\frac{2 \rho_p r_p^2 u_{\tau}^2}{9 \rho \nu^2}$ [-]}
\nomenclature[G]{$\varepsilon$}{turbulent kinetic energy dissipation rate [m$^2$ s$^{-3}$]}
\nomenclature[G]{$\eta$}{particle removal efficiency [\%]}
\nomenclature[G]{$\theta$}{exit flow angle [$^\circ$]}
\nomenclature[S]{$d$}{deposition}
\nomenclature[S]{$p$}{particle}
\nomenclature[S]{$b$}{blade tip}
\nomenclature[S]{$mid$}{blade mid-span}
\nomenclature[S]{$\infty$}{ambient}
\nomenclature[S]{$max$}{maximum}
\nomenclature[S]{$t$}{turbulent}
\printnomenclature
\end{framed}
\end{table*}


\section{Introduction}
\label{Sec:Intro}

\noindent The concentration of environmental pollutants, such as coarse and fine particle matter (PM$_{10}$ and PM$_{2.5}$), has increased at an alarming rate since pre-industrial levels. The most harmful particles are those with aerodynamic diameters below 2.5$\mathrm{\mu{m}}$ (PM$_{2.5}$) and are projected to increase in concentration up to ten-fold by 2050 in a business-as-usual scenario \cite{Pozzer2012}. Of course, the most serious concern is the impact this has on health and mortality rates \cite{Lelieveld2015}, however, the presence of particle pollutants also adversely affects the reliability of technologies by accumulating on surfaces \cite{Tencer2008}. The deposition of particle pollutants can lead to mechanical, chemical and electrical failures in electronic systems that are either located outdoor, or in an insufficiently controlled indoor environment. Deposition mechanisms can include particle diffusion, gravitational settling, thermophoresis, electrophoresis and photophoresis \cite{Tencer2008}. In forced cooling systems, however, advection is typically dominant over other transport mechanisms.

Axial flow fans are the common air cooling method in many engineering applications, from small scale portable electronic devices to much larger scale air conditioning systems for vehicles and buildings. The primary function can be to maintain the reliability of temperature-sensitive equipment or provide comfortable living conditions for occupants. This wide applicability has led to investigations on the aerodynamic performance characteristics, downstream flow distributions and the resulting global and local heat transfer performance \cite{Yen2006,Grimes2004,Stafford2010b}.  

Although the main flow is in the axial direction, fan rotation introduces radial and tangential velocity components, leading to a swirling exit flow that expands downstream. The presence of the fan hub adds additional complexity as air is expelled from an annular opening. Investigations on aerodynamic performance of three fan designs with different impeller angles and shrouding were performed by Yen and Lin \cite{Yen2006} using particle image velocimetry (PIV) and standardised bulk fan characterisation measurements. Non-uniform axial and radial velocity profiles were measured, with peak velocities near the blade tip. Vortices form in the region behind the hub, as the high speed flow interacts with stagnant air. Stafford \textit{et al.} \cite{Stafford2010b} used infrared thermography and a heated-thin foil technique to show that the radial averaged heat transport is also maximum in this region of peak velocity. Fluctuations in heat transfer coefficient were found to be an order of magnitude higher downstream of the fan hub, confirming the unsteadiness of the flow in this region and the effect it can have on transport phenomena. 

Yen and Lin \cite{Yen2006} also noted that shrouding around the fan provides the most stable exit flows compared to operating without a shroud, and the exit flow is less sensitive to changes in impeller angle. By adding winglet-blades, fan efficiency improved by approximately 10\% and was attributed to an increase in the lift-to-drag ratio. At the fan blade level, Estevadeordal \textit{et al.} \cite{Estevadeordal2000} examined dominant flow features around the pressure and suction sides by performing PIV measurements synchronised with the blade movement. Visualisation images revealed the unsteady flow generated in the blade passage including parallel wakes, axial streaks and separation. Yoon and Lee \cite{Yoon2004} showed that the periodic motion of the rotating blades translates to a periodic flow structure downstream when performing phase-averaging of the velocity fields. Stereoscopic PIV measurements indicated the largest out-of-plane velocities occur from the vortices shed by the blade tip.   

Most studies have provided insights by examining a fan operating in isolation, or in an idealised configuration. Grimes and Davies \cite{Grimes2004} investigated the relationship between air flow and heat transfer for an axial fan cooling a mock electronic system. `Push' and `pull' configurations were examined experimentally using PIV together with infrared thermography measurements of a printed circuit board with heating elements. For a fan sucking air through the system (pull), an ordered, uniform flow was apparent. For a fan blowing air through the system (push), however, the non-uniform and unsteady swirling flow improved thermal mixing, resulted in locally high incident flows, and considerably reduced the average component temperature by $16\mathrm{^{\circ}C}$ compared to a pull arrangement.   

Despite the widespread use of these air moving devices and increasing pollution levels, insights into the mechanisms of particle deposition have received limited attention. Indeed, previous research efforts have traditionally focused on applications such as gas and steam turbines \cite{Guha2008,Slater2003}, flue gas heat exchangers \cite{Tang2020}, respiratory tracts \cite{Guha2008} and pipe and ventilation duct flows among others \cite{Sippola2005,Wu2012}. In the latter, deposition rates in ventilation ducts are normally investigated in well-defined fully-developed and developing turbulent conditions. As evidenced by the literature, flows produced by air moving devices are inhomogeneous with complex spatio-temporal structures. While the pressure, flow and power coefficients are sensitive to fan geometry \cite{Bleier1998}, the previous literature shows that a swirling diverging flow structure is synonymous with axial fan operation. This general flow feature also produces similar convection heat transfer patterns for different axial fan designs in inertia dominant regimes ($Re > 2000$) \cite{Stafford2010b,Stafford2010a}. 

As environmental pollution has increased, the deposition of particle matter has become a primary concern for equipment reliability in forced-air cooling systems. A detailed understanding of the relationship between the complex flow fields generated by rotating fans and the accumulation of particle matter is required to improve mitigation strategies. Furthermore, accurate predictions of particle deposition also require the correct treatment of near-wall turbulence \cite{Zhang2009}, signifying the prerequisite role of highly resolved experimental data for model validation. In this work, instantaneous and ensemble-average velocity fields of representative PM$_{2.5}$ particles are measured using particle image velocimetry. This investigation is combined with observations of long-term particle deposition inside fan-cooled equipment, to develop an understanding on the relationship between the unsteady turbulent flow and spatial variability of deposition rates.     

\section{A description of particle transport and deposition}
\label{Sec:Theory}
\noindent Particles can be transported from the fluid onto a surface by a number of different mechanisms discussed above \cite{Tencer2008,Guha2008}. The mass transfer rate, $J$, of particles to a surface can be described as: 

\begin{equation}
\label{eq:J}
J = U_d c
\end{equation}

\noindent where $U_d$ is the deposition velocity and $c$ is the bulk concentration in the flow. Treating the particles as a hypothetical ideal gas, the motion can be described by Eulerian conservation equations for continuity and momentum. Performing Reynolds-averaging and simplification by neglecting Reynolds stress terms in the streamwise direction and triple correlations of primed quantities, Guhu \cite{Guha1997} showed that the particle flux in the wall-normal direction for fully-developed flow can be expressed as:  

\begin{equation}
\label{eq:Jterms}
J = \overline{\rho}_p \overline{v}_p + \overline{\rho'_p v'_p} = -(D + D_t) \frac{\partial\overline{\rho}_p}{\partial{y}} - D_T \overline{\rho}_p \frac{\partial\ln T}{\partial{y}} + \overline{\rho}_p \overline{v}^{c}_{p}
\end{equation}

\noindent Here, the absolute velocity terms have been decomposed into diffusive and convective parts, with the particle mass flux due to turbulent fluctuations modelled by gradient diffusion $\overline{\rho'_p v'_p} = -D_t (\partial\overline{\rho}_p/\partial{y})$. This explicitly shows a number of mechanisms that can contribute to mass transfer, including molecular and turbulent diffusion, thermophoresis, eddy impaction and particle inertia. The general Eulerian description conveniently allows for the inclusion of relevant terms and other forces that may be relevant to a particular mass transport problem \cite{Guha2008}.   

Many forced-air cooled systems, such as that investigated in the present study, involve advection-dominated turbulent particle mass transport with $10^2 < Pe < 10^5$. In our investigation on fine PM$_{2.5}$ particulates, the rate of Brownian diffusion is small compared with the transport of fluid momentum with $10^2 < Sc < 10^6$. Considering $Sc_t \approx 1$ in turbulent flows \cite{Shimada1993}, turbulent diffusion has a significantly greater contribution to particle deposition. The deposition regime examined in this experimental work spans turbulent diffusion and eddy impaction with $0.1 < \tau^+ < 1$.      


\section{Experimentation}
\label{Sec:Exp}

\subsection{Observations of particle accumulation}

\begin{figure}[p]
\centering\includegraphics[width=0.95\linewidth]{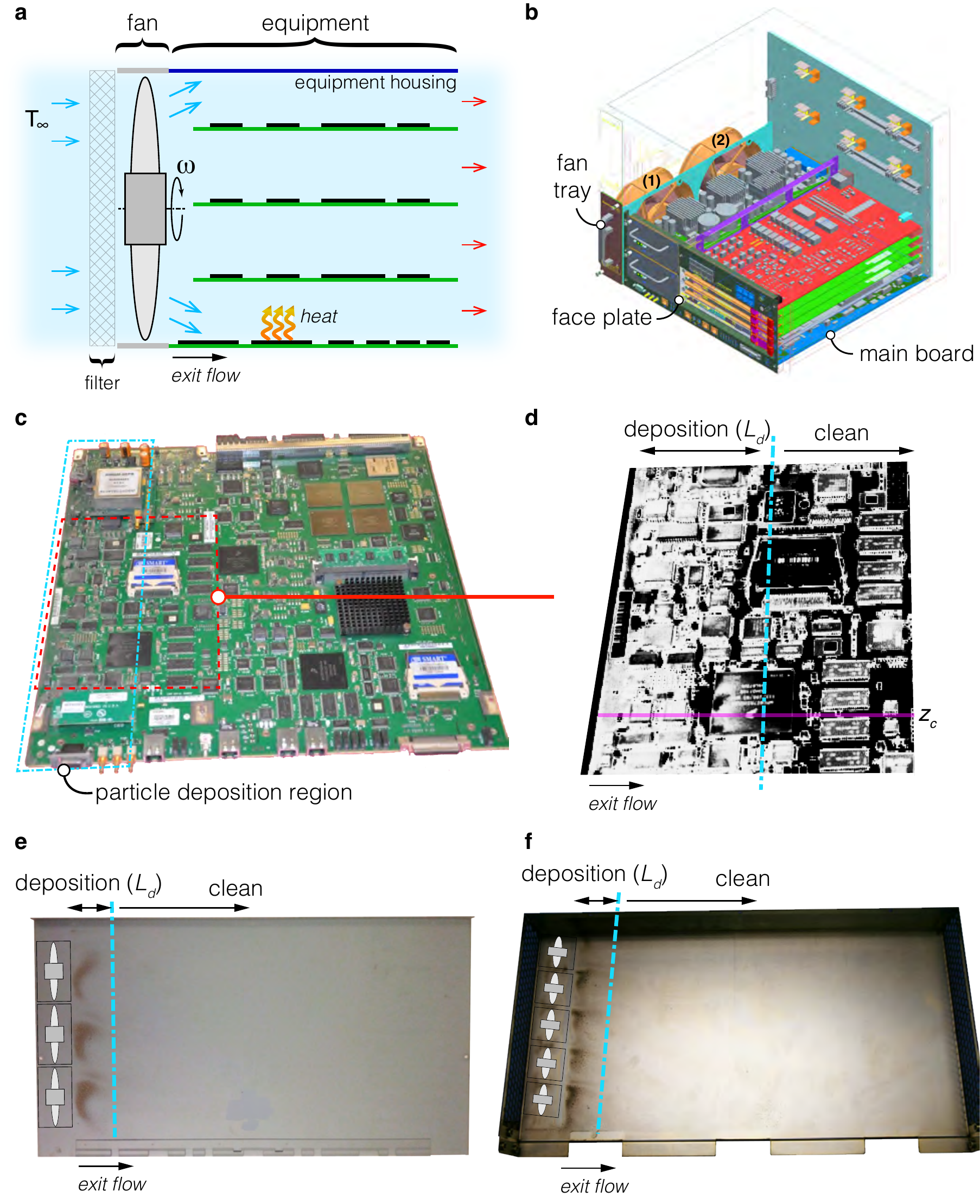}
\caption{Accumulation of particle pollution in fan-cooled equipment. a) Schematic of the typical `push' cooling arrangement. b) A forced-air cooled telecommunications base station together with c,d) field observations of deposition on the main board. Also shown are field observations of deposition on the equipment housings of different e) three fan and f) five fan cooled service aggregation routers.}\label{fig:9916-dust}
\end{figure}

\noindent An illustration of forced-air cooling, with fans operating in a typical `push' configuration, is shown in Fig. \ref{fig:9916-dust}a.  Visual assessments on various indoor forced-air cooled electronic equipment operating continuously in the field for $\approx {1-2}$ years were obtained. This equipment was carefully disassembled to gain optical access to areas of enhanced particle deposition. Inspections of a telecommunications base station and service aggregation routers are provided in Fig. \ref{fig:9916-dust}b-f. Regions of elevated levels of particle deposition were observed downstream of cooling fans on circuit boards (Fig. \ref{fig:9916-dust}c,d) and equipment housings (Fig. \ref{fig:9916-dust}e,f). This occurred within approximately one fan diameter of the exit flow. While each equipment has a similar general architecture (Fig. \ref{fig:9916-dust}a), the electronic system layout, dimensions, number of axial fans, and axial fan design and performance are all dissimilar. As different equipment and axial fans produced the same qualitative deposition features, one test case was chosen to study the general mechanisms behind this localized preferential particle deposition process. This base station test case equipment is shown in Figure \ref{fig:9916-dust}b-d. A fan tray containing two axial fans ($r_b = 72$ mm, $400 < \omega < 500$ rad s$^{-1}$, $Q_{max} = 0.164$ m$^{3}$ s$^{-1}$, $\Delta{p_{max}} = 370$ Pa) control the component temperatures within the equipment. Deposition occurred on the main board over a length $L_d$ (Fig. \ref{fig:9916-dust} c,d) and adjacent to the cooling air flow exiting the fan tray (Fig. \ref{fig:9916-dust} a,d). Localized preferential deposition was also observed on the top cover of the equipment housing, similar to that shown in Fig. \ref{fig:9916-dust}e,f. Similar deposition patterns exist on both the heat generating printed circuit board ($\Delta{T_{max}} = T - T_{\infty} \approx 50$ K) and equipment covers ($\Delta{T} \approx 0$), suggesting that thermophoresis has a negligible role on the particle flux compared to the turbulent flow contributions (Eq. \ref{eq:Jterms}). 

\subsection{Velocity measurements}\label{sec:piv}

\begin{figure}[t]
\centering\includegraphics[width=1\linewidth]{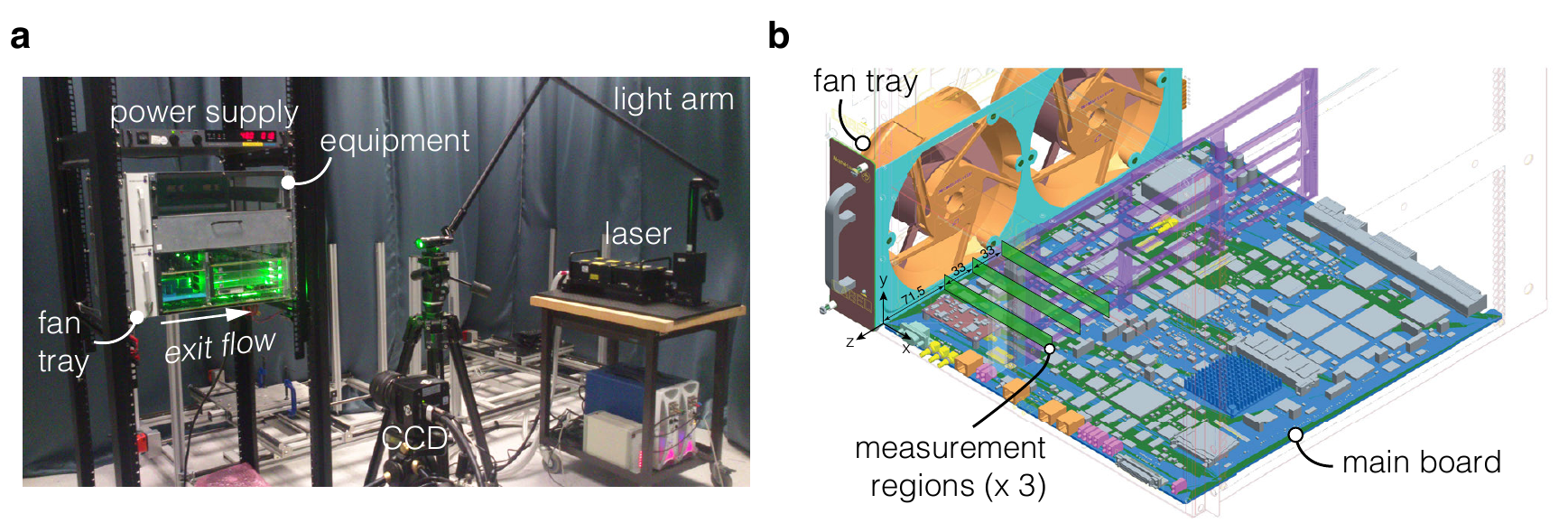}
\caption{\textit{In situ} particle image velocimetry on fan cooled equipment. a) Experimental arrangement for the equipment shown in Fig.\ref{fig:9916-dust} b-d including b) the location of the velocity measurement planes (internal circuit boards removed for illustration purposes only).}\label{fig:PIV-setup}
\end{figure}

\noindent The qualitative field observations suggest that the fluid motion from the exit flow is strongly linked to the particle deposition. This was investigated by performing velocity field measurements \textit{in situ} on the fan-cooled equipment around this area of interest using particle image velocimetry (PIV). The experimental arrangement is shown in Fig. \ref{fig:PIV-setup}a. The equipment was rack-mounted along with an external power supply that controlled the rotational speed of both axial fans. This arrangement was contained within a dedicated testing environment approximately 8 m (L) $\times$ 3 m (W) $\times$ 3 m (H). Minor modifications to the equipment face plate (Fig. \ref{fig:9916-dust}b) were necessary to gain optical access for recording velocity field data near the deposition region (Fig. \ref{fig:PIV-setup}a). This modification was performed with negligible effect on the flow behaviour in the measurement region. Three measurement planes outlined in Fig. \ref{fig:PIV-setup}b (green planes) were examined. The middle measurement region was located in a ${x-y}$ plane aligned to the central axis of the fan labelled (1) in Fig. \ref{fig:9916-dust} b. This reference location is at $z = z_c$. The two other measurement regions were positioned relative to this middle measurement plane at distances $z \approx z_c + 0.5r_b$ and $z \approx z_c - 0.5r_b$, respectively.

The PIV system consisted of a laser, light arm, CCD camera and synchroniser. A thin laser sheet ($\approx 0.5$ mm) with 532 nm wavelength was generated using a Quantel Big Sky Laser. A light arm was connected to the laser to allow the positioning of the laser system away from the air flow outlet of the equipment, thus avoiding any flow restrictions (Fig. \ref{fig:PIV-setup}a). The laser sheet illuminated glycol-based tracer particles ($d_p < 2.5 \mathrm{\mu{m}}$) twice in succession for a two-dimensional measurement plane (Fig. \ref{fig:PIV-setup}b). The tracer particles were introduced into the test environment using a Magnum 2500 hazer and JEM Pro Haze Fluid (relative density of 1.05 at $20\mathrm{^{\circ}C}$). A homogeneous particle distribution was obtained after approximately 10 min of continuous issuing of tracer particles. Once sufficient tracer particle density was achieved, the hazer was turned off for 1 min prior to recording velocity field data. A Dylos DC1700 air quality monitor was used to estimate the particle size that was generated from the hazer. This provided a measurement for PM$_{2.5}$ ($d_p < 2.5 \mathrm{\mu{m}}$) and PM$_{10}$ ($2.5 \mathrm{\mu{m}} < d_p < 10 \mathrm{\mu{m}}$), respectively. Over 90\% of the particles sensed were in the range PM$_{2.5}$. Apparent densities of environmental PM$_{2.5}$ vary depending on global location, pollution source and also diurnal and seasonal conditions \cite{Liu2015}. The composition and relative density of the tracer particles suggest that their apparent density would also be lower than that observed for environmental PM$_{2.5}$, which can range $1.3 - 1.8$ g cm$^{-3}$ \cite{Liu2015}. Estimating a difference in apparent density of $\approx 20-70\%$ and comparing the particle response time to the turbulent Kolmogorov time scale \cite{Dou2018}, the inertial regimes for both environmental PM$_{2.5}$ pollutants and the tracer particles are found to be equivalent with a particle Stokes number, $St \approx 0.1-0.17$. It is reasonable to assume, therefore, that velocity fields obtained using the tracer particles in the present study are representative of the flow of environmental PM$_{2.5}$ pollutants.        




A TSi PowerView Plus 4MP CCD camera was positioned perpendicular to the laser sheet plane to record the scattered light from the particles on two separate frames. The camera and laser were synchronized using a TSi laser pulse synchronizer. Each image pair was split into interrogation regions that were 32 $\times$ 32 pixels and processed using TSi Insight 4G software. Timing between laser pulses was adjusted based on fluid velocities in the region of interest. The timing was selected to ensure that the relative image displacement was kept below or equal to 0.25 for each interrogation region, as recommended by Keane and Adrian \cite{Keane1990}. For the velocity data presented in this paper, the resultant timing set between pulses ranged $\mathrm{3 \mu{s} < \Delta{t} < 30 \mu{s}}$. Velocity vector fields presented in this paper have a spatial resolution of 160 $\mathrm{\mu{m}}$ and were acquired at a frequency of 7.5 Hz. Hence, these measurements represent the two-component velocity ($u_p,v_p$) of a group of $\mathrm{PM}_{2.5}$ particles over a 160 $\mathrm{\mu{m^2}}$ region.

The PIV technique produced velocity field data for PM$_{2.5}$ tracer particles in a Eulerian scheme, with $\mathbf{U}=\mathbf{U}_p$. This is compatible with the theoretical description of advection-diffusion particle transport introduced in Section \ref{Sec:Theory}. Velocity field data was decomposed into $\mathbf{U}  = \overline{\mathbf{U}} + \mathbf{U}'$, where $\mathbf{U}$ consists of streamwise and wall-normal velocity components, $(u,v)$. In the context of particle mass flux (Eq. \ref{eq:Jterms}), these measurements are absolute velocities that include the combined effects that contribute to particle transport. This decomposition into mean and fluctuating velocities was used to examine the ensemble-average flow field and also to perform a quadrant analysis \cite{Wallace2016}. The vast majority of literature works formally from time-averaged quantities in the description of particle mass and momentum conservation equations \cite{Wu2012}. This work also investigates the turbulent events that promote wall-normal momentum and inward-outward particle transport. The quadrant analysis classified the products of the velocity fluctuations into the categories Q1 $(+u',+v')$, Q2 $(-u',+v')$, Q3 $(-u',-v')$ and Q4 $(+u',-v')$. This classification, more commonly used for turbulent boundary layer studies, has been exploited herein to reveal details on particle \textit{deposition events}, observed from $-v'$ quadrants Q3 and Q4.

\subsection{Velocity measurements on filtered flows}

\begin{figure}[t]
\centering\includegraphics[width=0.5\linewidth]{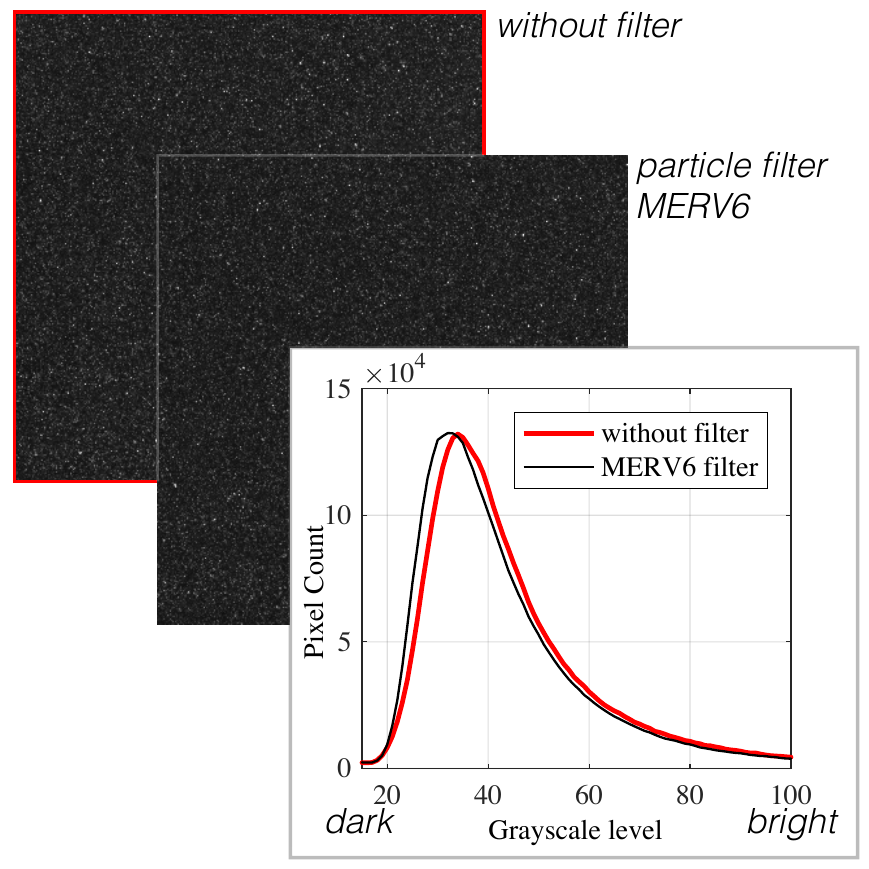}
\caption{Particle images and image histograms for the baseline case without a filter and with a MERV 6 filter for PM$_{10}$ removal.}\label{fig:filter-exp}
\end{figure}

\noindent Particle filtering is the common approach for removing particulate matter before it can enter the interior of buildings and equipment. Filtering of large PM$_{10}$ pollutants is often achievable, however, high-efficiency filtering of PM$_{2.5}$ is normally prohibited in the majority of applications due to the adverse effects it has on cooling air flow rate and thermal performance. Velocity measurements were performed on filtered air flows to investigate the effect of PM$_{10}$ filtering on the flow behaviour of PM$_{2.5}$. Two different filters (Minimum Efficiency Reporting Value, MERV 5 and MERV 6 \cite{ASHRAE}) were installed at the air inlet to the fans (Fig. \ref{fig:9916-dust}a). The MERV 5 filter has a removal efficiency of 20-35\% for $3\mathrm{\mu{m}} < d_p < 10 \mathrm{\mu{m}}$. The MERV 6 filter has a higher removal efficiency of 35-50\% for $3\mathrm{\mu{m}} < d_p < 10 \mathrm{\mu{m}}$. Particle images from within the equipment are shown in Fig. \ref{fig:filter-exp} for the cases of without a filter and with MERV 6 filtering. Histograms of the pixel intensities indicate the dark and bright levels across the image, and have been used here to describe changes in the concentration of tracer particles (i.e. a darker image is equivalent to a lower concentration of tracer particles). Compared to the histogram obtained for the baseline case without a filter, the addition of a MERV 6 filter has only a minor effect on PM$_{2.5}$ concentration in the equipment. Hence, a similar analysis of the flow fields using the PIV technique described above was also implemented for the PM$_{10}$ filtered air flows considered here.             

\subsection{Uncertainty}

\noindent Fixed errors for the PIV system were estimated to produce a relative uncertainty in a velocity magnitude of $< 2\%$. 
Deposition rate has a linear dependence with particle concentration (Eq. \ref{eq:J}), and a suitable concentration of PM$_{2.5}$ tracer particles was chosen for accurate velocity measurements. The particle image density (NI) was $17 < \mathrm{NI} < 25$ per interrogation region, ensuring that a high probability of valid detection rate was achieved ($> 95\%$) \cite{Keane1990}. This choice of particle image density also lowered the measurement uncertainty in particle image shift as recommended by Raffel \textit{et al.} \cite{Raffel2007}. This assessment was based on an average particle concentration across all interrogation regions of the image, and any local differences in particle concentration were deemed negligible by maintaining uniform seeding density. A minimum of 1000 image pairs were recorded for each measurement location and the convergence of full field measurement statistics was considered \cite{Stafford2012a}. The ensemble average velocity field sufficiently converged toward the time average within 5\% using this sample size. 

\section{Results and discussion}
\label{Sec:Results}
\begin{figure}[p]
\centering\includegraphics[width=0.5\linewidth]{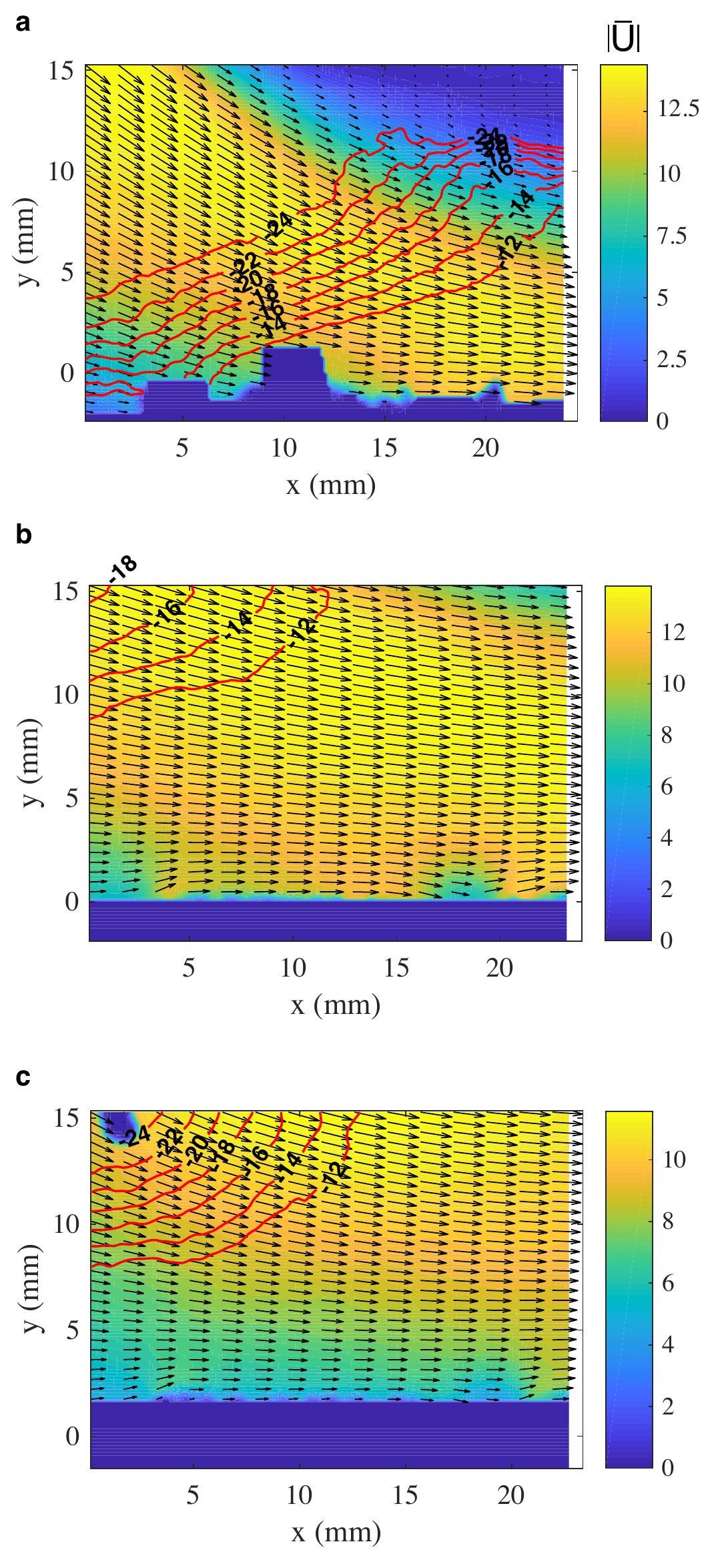}
\caption{Ensemble-averaged velocity fields in the exit flow region with measurement planes located at a) $z \approx z_c - 0.5r_b$, b) $z = z_c$  and c) $z \approx z_c + 0.5r_b$. Isocontours of $\theta = \tan^{-1}\left(\overline{v}/\overline{u}\right)$ included for $-12^\circ < \theta < -24^\circ$. Operating speed $\omega = 419$ rad/s. }\label{fig:piv-3-planes}
\end{figure}

\noindent The ensemble-average velocity fields in the exit flow region and across the three measurement planes (Fig. \ref{fig:PIV-setup}b) are shown in Fig. \ref{fig:piv-3-planes}. Differences in the surface topography in Fig. \ref{fig:piv-3-planes}a are due to the layout of electronic components on the main board (Fig. \ref{fig:9916-dust}c,d). Despite this, together with the three-dimensional behaviour of rotating fan flows, the qualitative trends in the streamwise and wall-normal directions ($x,y$) are consistent for each plane along the spanwise direction, with flow impinging onto the surface at a shallow angle. This angle is shown to vary typically 18-24$^\circ$ beneath the horizontal and depending on the position of the plane in the $z$-direction.  The flow angle is produced by the radial velocity component introduced from the fan rotation, and is illustrated with a simple sketch in Fig. \ref{fig:9916-dust}a. This qualitative similarity across all three planes agrees with the field observations of enhanced deposition in Fig. \ref{fig:piv-3-planes}c-d. That is, in addition to these similarities in the velocity fields, the deposition length ($L_d$) is also consistent along the main board where the air flow exits. A single centrally located plane at $z = z_c$, therefore, has been chosen for the following detailed analyses on mean flow fields and turbulent particle deposition events. 

\subsection{Mean flow deposition mechanism}

\begin{figure}[htb!]
\centering\includegraphics[width=1\linewidth]{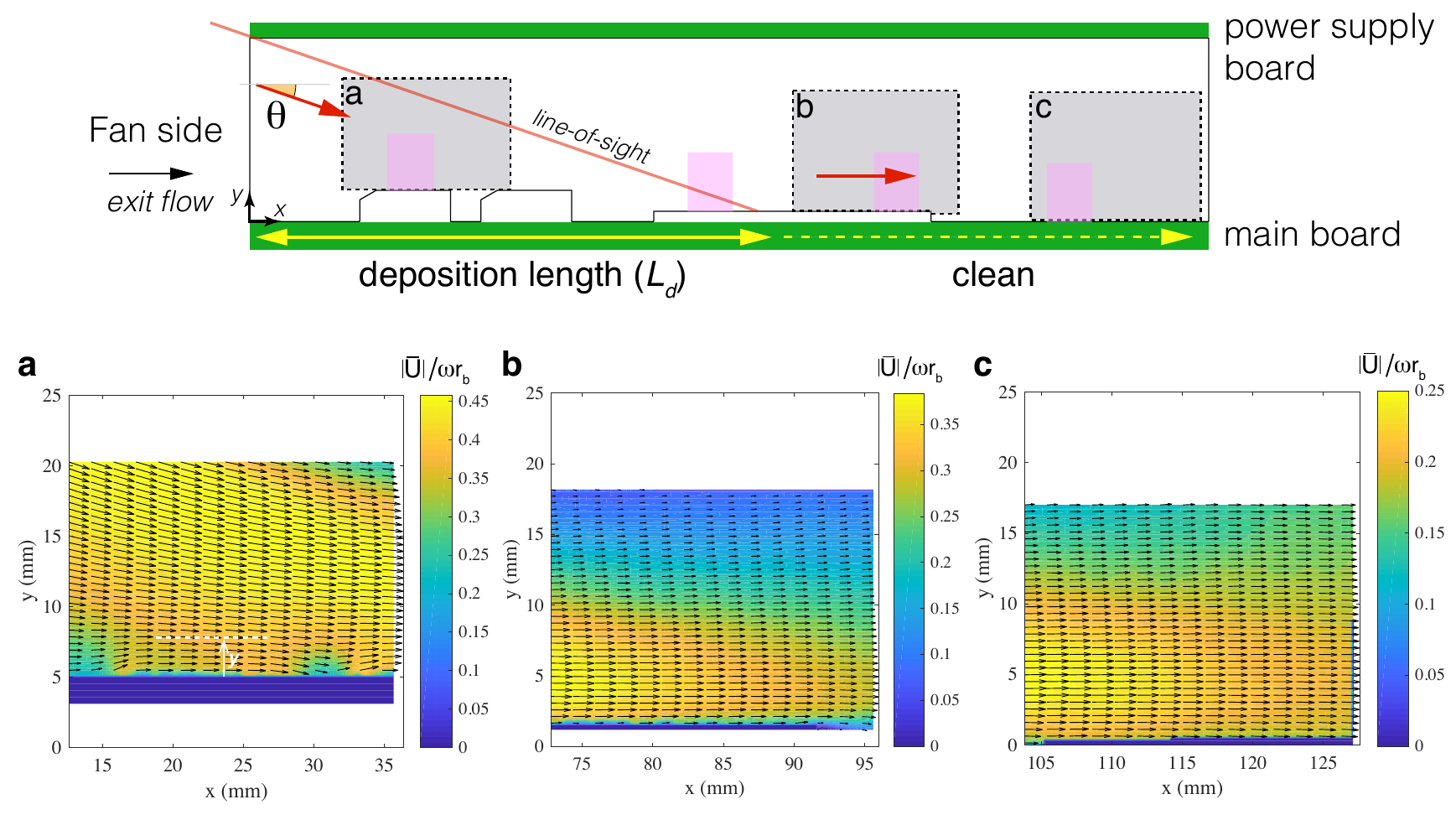}
\caption{Ensemble-averaged velocity fields above particle deposition and `clean' surfaces. A schematic of the measurement plane at $z = z_c$ indicating the location of velocity field regions (a-c) relative to the deposition observations in Fig. \ref{fig:9916-dust}d. Operating speed $\omega = 419$ rad/s.}\label{fig:piv-centre-1-4-6}
\end{figure}

\noindent A wider view of the mean flow above particle deposition and clean surfaces is shown in Fig. \ref{fig:piv-centre-1-4-6}. The term `clean' is used loosely here to distinguish surfaces from those with substantial particle deposition (Fig. \ref{fig:9916-dust}d). The flow initially approaches the surface at an angle $\theta \approx -19^\circ$ (taken as the angle in the upper left corner of the measurement region, Fig. \ref{fig:piv-3-planes}b). At $x \approx r_b$ downstream, the flow is fully redirected to the streamwise direction ($\theta = 0^\circ$) when considered as a time-average. Taking a line-of-sight at the angle of the exit flow, and within the confines of the channel formed between power supply and main boards, an intersection occurs at the transition from deposition to clean surfaces (see schematic in Fig. \ref{fig:piv-centre-1-4-6}, Fig. \ref{fig:9916-dust}d). This high velocity oblique flow promotes particle mass transport to the surface. The maximum velocity decreases from $|\overline{U}|/\omega{r_b} \approx 0.45$ to 0.25 from the immediate exit flow to the clean downstream region at $x \approx 1.5r_b$.  Interestingly, the redirection of the flow after impingement has less of an effect on the shear velocity. This reduces by only $\approx 10\%$ from the deposition region in Fig. \ref{fig:piv-centre-1-4-6}a to the transition to a clean surface in Fig. \ref{fig:piv-centre-1-4-6}b. This is due to the flow maintaining a high velocity close to the wall ($y < 10$ mm) immediately after impingement as it forms a wall jet zone. Further downstream, the shear velocity reduces by $\approx 20\%$ as the wall jet expands into the flow channel and the near wall gradients reduce (Fig. \ref{fig:piv-centre-1-4-6}c). 
At this location downstream ($x \approx 1.5r_b$), the deposition velocity for $d_p = 2.5 \mu$m particles is reduced by $\approx 30\%$. The significance of operating in the eddy diffusion impaction regime ($0.1<\tau^+<1$), where substantial changes in deposition velocity occur \cite{Guha2008}, is revealed when considering that coarser particles can enter air cooling systems through either unfiltered or filtered environments with low removal efficiency. For coarser particles with $d_p < 10 \mu$m, the deposition velocity reduces by an order of magnitude or more in the downstream clean region. 

\begin{figure}[htb!]
\centering\includegraphics[width=0.5\linewidth]{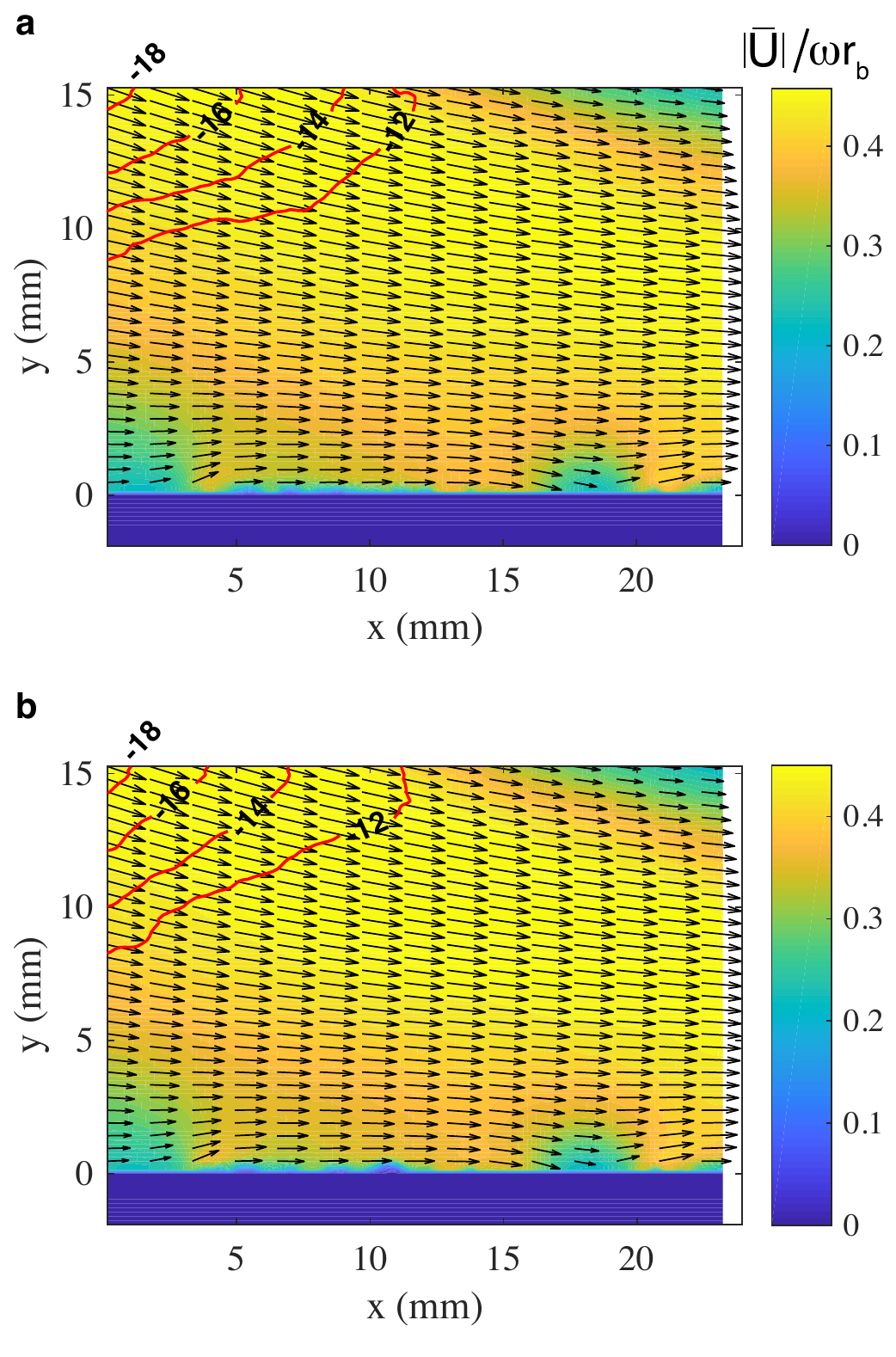}
\caption{Ensemble-averaged velocity fields and $\theta$ isocontours for a) $\omega = 419$ rad/s and b) $\omega = 497$ rad/s at $z = z_c$. Operating speed $\omega = 419$ rad/s.}\label{fig:piv-2-speeds}
\end{figure}

The effect of rotational speed on the mean particle flow behaviour and deposition pattern can be generalised. Ensemble-averaged exit flow fields scale with rotational speed, as shown in Fig. \ref{fig:piv-2-speeds}. This occurs as volume flow rate from axial fans vary linearly with speed ($Q \sim \omega$) at high Reynolds numbers. By normalising the local velocity field with fan blade tip speed ($\omega r_b$), the normalised velocity magnitude and impingement angle are shown to be invariable with rotational speed. This suggests that the preferential deposition length scale , $L_d$, will be similar for different rotational speeds around the nominal fan speed. The mass transfer rate (Eq. \ref{eq:Jterms}), however, will change as the absolute velocity and turbulent fluctuations are influenced by a change in fan speed. The particle relaxation time is modified, and ultimately the deposition velocity, $U_d$. Although the deposition pattern is likely to be similar, this alters the quantity of particles deposited on the surface over time. If the direction of this speed change is towards higher $\omega$, the detrimental accumulation of particle mass on sensitive components will be accelerated.

\subsection{Particle deposition events}

\begin{figure}[htb!]
\centering\includegraphics[width=1\linewidth]{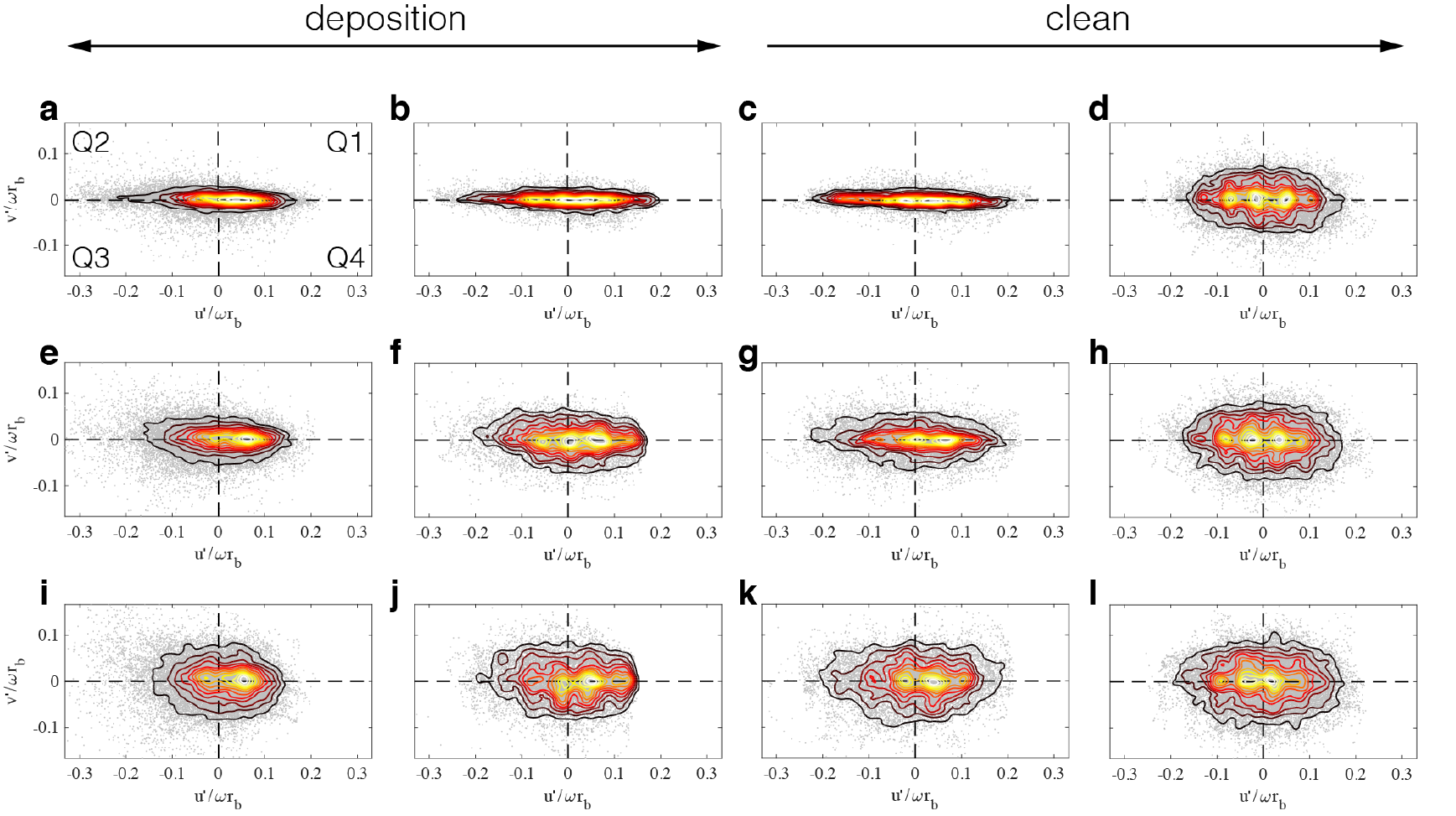}
\caption{Velocity fluctuations ($u',v'$) and joint probability density functions ($P(u',v')$) for deposition and clean regions at wall distances a-d) $y^+ = 14,11,10,11$, e-f) $y^+ = 28,28,26,28$ and i-l) $y^+ = 49,50,47,49$. The shear velocities ($u_{\tau}$) are a,e,i) 0.532 m/s, b,f,j) 0.518 m/s, c,g,k) 0.480 m/s, and d,h,l) 0.425 m/s. Operating speed $\omega = 419$ rad/s.}\label{fig:quad-yplus-10-28}
\end{figure}

\noindent The mean flow characteristics offer a partial insight to the particle deposition observations. These flows are stochastic by nature, and to reveal the physical mechanisms and events that promote deposition, investigation of the time-varying statistics was performed using a quadrant analysis on the fluctuating components of velocity ($u',v'$) at various wall distances $10 < y^+ < 50$ (Section \ref{sec:piv}). The velocity fluctuations and joint probability distribution functions ($P(u',v')$) are shown in Fig. \ref{fig:quad-yplus-10-28} for four sample zones located along the deposition and clean surface regions. These are indicated on the schematic in Fig. \ref{fig:piv-centre-1-4-6} (pink boxes). The gray points are PIV measurements, with contours of $P(u',v')$ superimposed. The color levels on the contour plots describe the density of measurements, with bright regions indicating the highest density of $u'$ and $v'$ fluctuations. Quadrants Q1-Q4 have been labelled in Fig. \ref{fig:quad-yplus-10-28}a and describe `outward' interaction ($+u',+v'$), ejection ($-u',+v'$), `inward' interaction ($-u',-v'$) and sweep events ($+u',-v'$). The majority of the joint probability distribution functions have an elliptical shape with a slightly negative correlation coefficient $\sim -0.1$ and are inclined in the ejection-sweep direction, a characteristic of turbulent boundary layer flows at similar wall distances as investigated in this work \cite{Wallace1977}. Moving further away from the wall, $P(u',v')$ becomes more rounded with $v'$ fluctuations growing as $y^+$ approaches 50. The furthest downstream region ($x \approx 1.5r_b$) is less negatively correlated and closer to isotropic turbulence with stretching in $u'$, as the wall jet expands and the flow becomes evenly distributed across the channel gap (Fig. \ref{fig:piv-centre-1-4-6}c). 

Within the deposition region, the ejection-sweep events had $\approx 5\%$ higher contribution to the overall Reynolds stress ($\overline{u'v'}$) than in the clean region for $y^+ \approx 10$. These quadrants (Q2,Q4) represent the vertical momentum fluxes, and subsequently promote the transport of particles away and towards the surface. Significantly, $\overline{u'v'}$ increased sixfold in the particle deposition region compared to the clean region, with the reduction in turbulent stresses occurring across the transition to a clean surface. This turbulent term is incorporated in the Eulerian description of mass transport through turbulent diffusion, $\overline{u'v'} \sim -D_t(\partial \overline{u} / \partial y)$ \cite{Guha1997}. The combination of increased turbulent stresses and shear velocity produced by the oblique impinging exit flow is a significant driver for enhancing deposition of particle pollutants in forced-cooled systems.       
 
\begin{figure}[htb!]
\centering\includegraphics[width=0.5\linewidth]{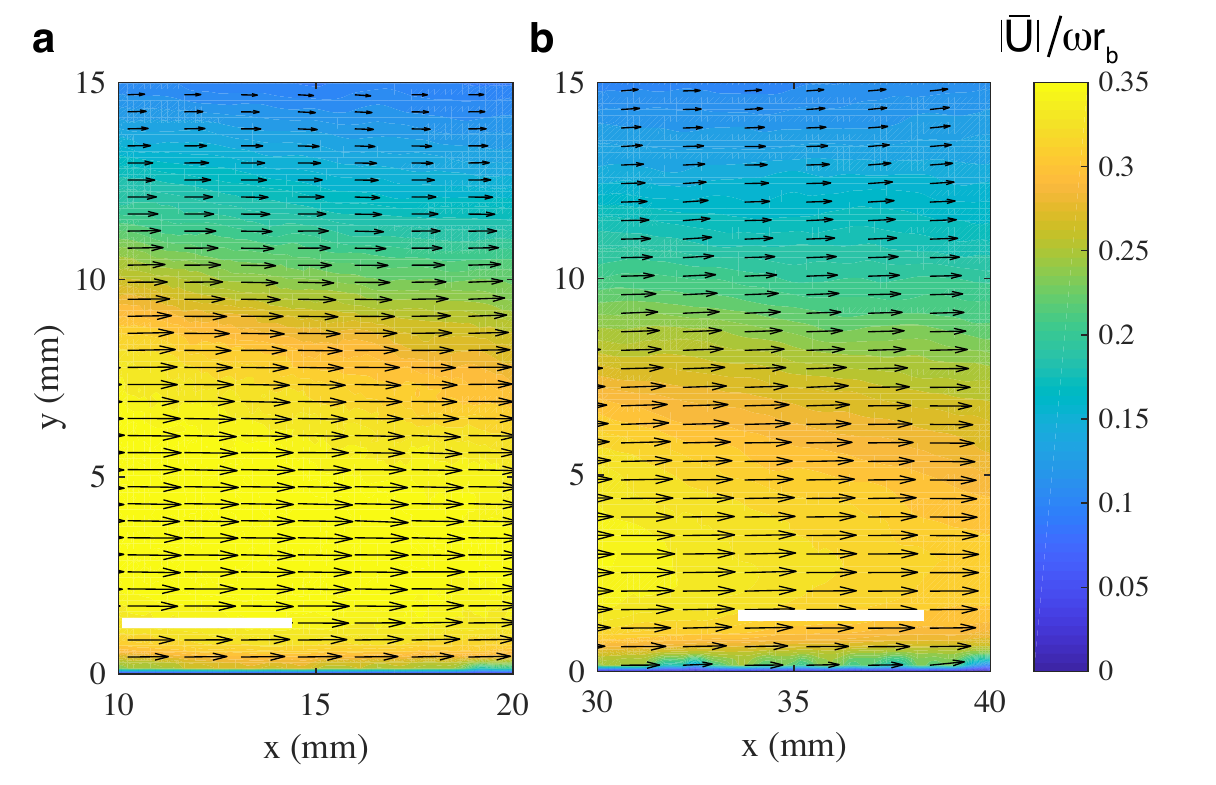}
\caption{Ensemble-average velocity fields across transition from a) deposition to b) clean surfaces. Locations of $y^+$ indicated by white line for $y^+ \approx 50$. Operating speed $\omega = 419$ rad/s.}\label{fig:dep-clean-avg}
\end{figure}

\begin{figure}[htb!]
\centering\includegraphics[width=0.5\linewidth]{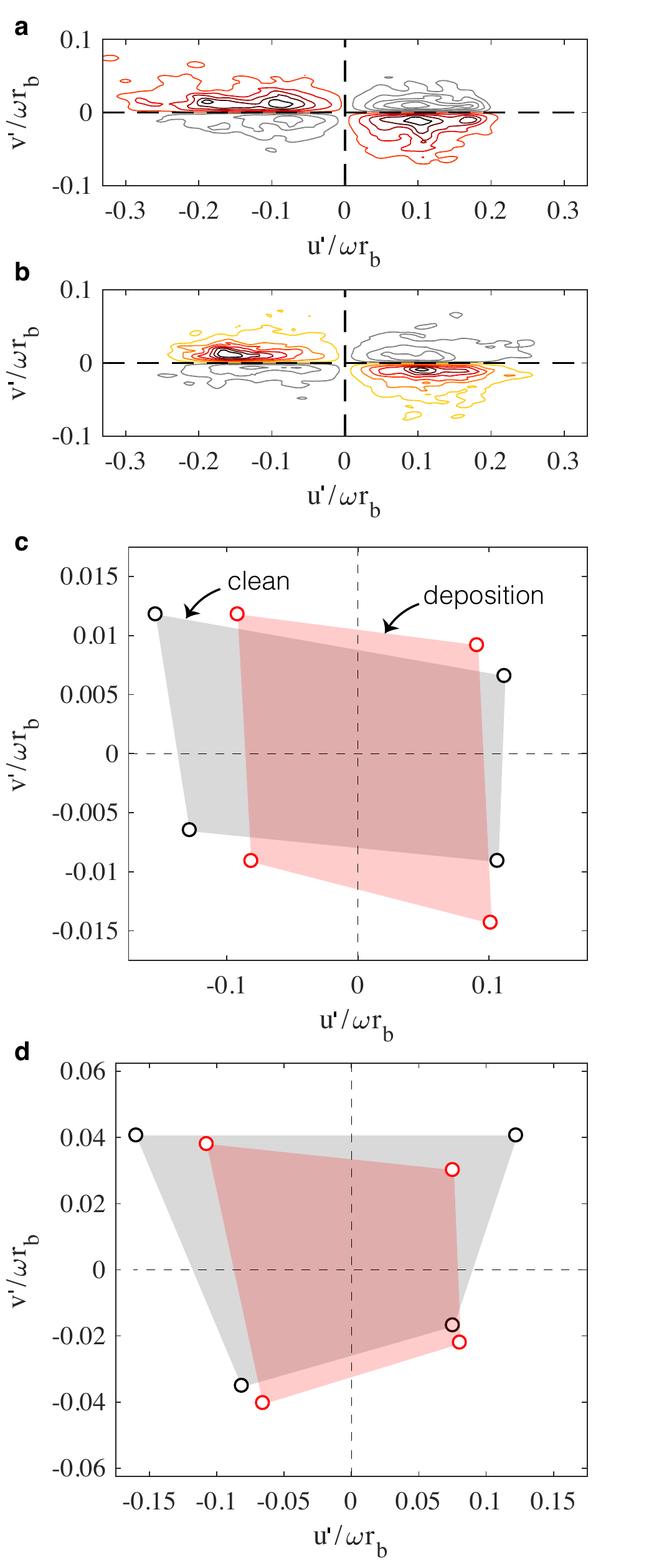}
\caption{Covariance integrand $u'v'P(u',v')$ for a) deposition $y^+ \approx 10$ and b) clean $y^+ \approx 11$ regions. Peaks in $u'v'P(u',v')$ for quadrants Q1-Q4 and wall distances c)  $y^+ \approx 10$ and d) $y^+ \approx 30$. Operating speed $\omega = 419$ rad/s.}\label{fig:CoVar-yplus-10-28}
\end{figure}

\begin{figure}[p]
\centering\includegraphics[width=0.85\linewidth]{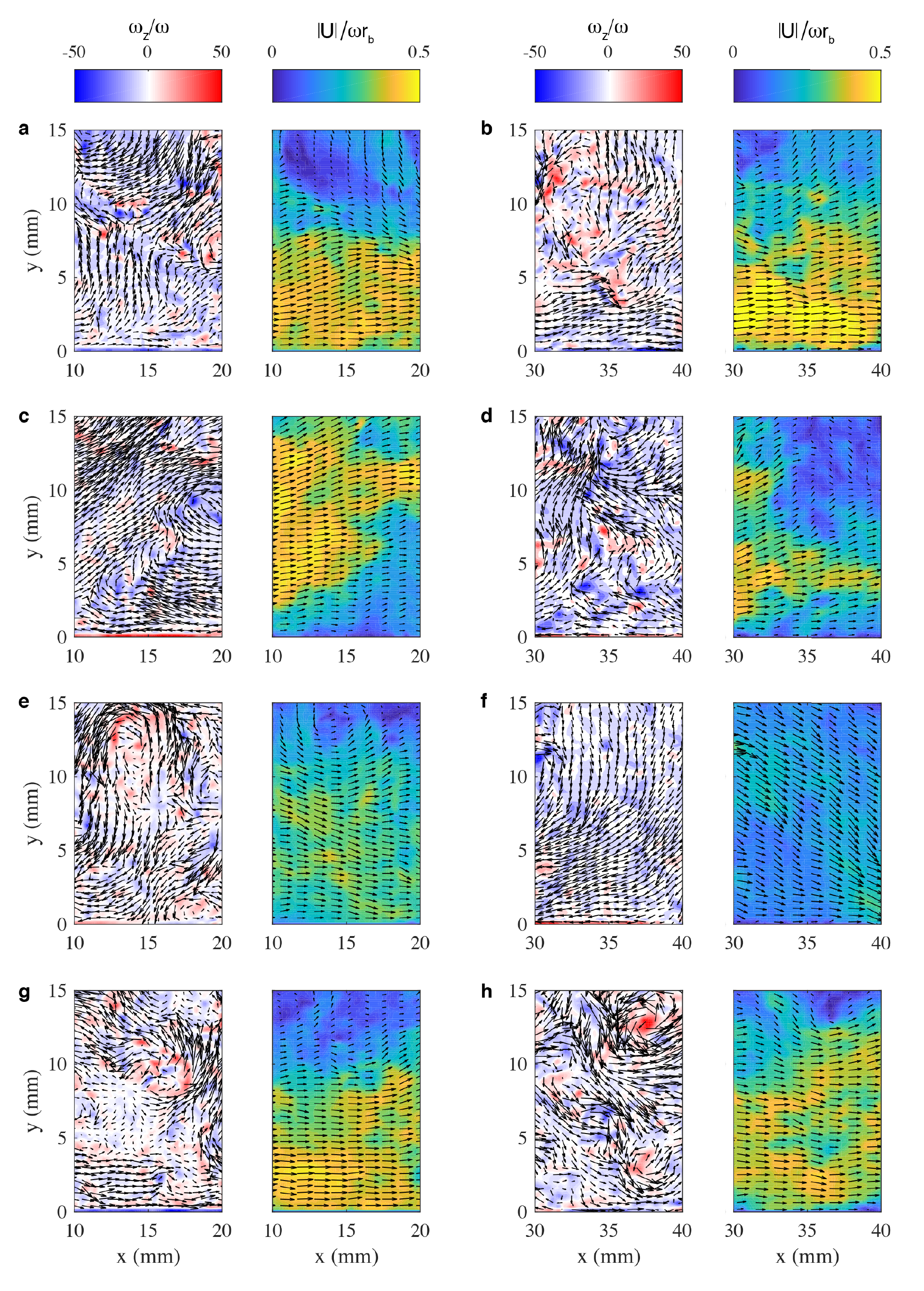}
\caption{Instantaneous vorticity fields and velocity fields selected at peaks in $u'v'P(u',v')$ (Fig. \ref{fig:CoVar-yplus-10-28}c) and for quadrants a,b) Q1, c,d) Q2, e,f) Q3 and g,h) Q4. Enhanced particle deposition observed on surface ($y = 0$) in a,c,e,g). Operating speed $\omega = 419$ rad/s.}\label{fig:vorticity}
\end{figure}

Changes in the ensemble-average flow field across the transition from particle deposition to clean surfaces is shown in Fig. \ref{fig:dep-clean-avg}. The larger Q2-Q4 contributions for deposition appear through covariance integrand ($u'v'P(u',v')$) plots in Fig. \ref{fig:CoVar-yplus-10-28}a-b. Turbulent stresses are spread over greater quadrant areas (Q2-Q4) suggesting more frequent occurrences of extreme ejection, inward interaction and sweeping events. Peaks in $u'v'P(u',v')$ have been selected and plotted in Fig. \ref{fig:CoVar-yplus-10-28}c-d for two wall distances. The location of these contour peaks are statistically important conditions for turbulent diffusion and eddy impaction mechanisms, representing the pairs of streamwise and wall-normal velocity fluctuations that most contribute to the Reynolds stress \cite{Wallace2016}. Instantaneous vorticity ($\nabla \times {\mathbf{U'}}$) and velocity flow fields that meet this streamwise and wall-normal pairing criteria for Q1-Q4 were selected at random from the velocity measurements and have been presented in Fig. \ref{fig:vorticity}. Small islands of positive (clockwise) and negative (anti-clockwise) vorticity increase in size and magnitude further away from the wall ($y=0$), as turbulent eddies support the transport of mass between the bulk flow and boundary layer.

Increases to $-v'$ of up to 65\% in the deposition region were observed at $y^+ \approx 10$ and 30, promoting the transport of particles towards the wall and deposition onto the surface (Fig. \ref{fig:quad-yplus-10-28}c-d). Streamwise $+u'$ levels are similar, indicating that sweeping of particles towards the surface is more dominant in the deposition region. This effect is shown in the local flow fields for $y<5$mm and Q4 events in Fig. \ref{fig:vorticity}g-h. Interestingly, streamwise flow deceleration ($-u'$, Q2-Q3) is lower in the deposition region (Fig. \ref{fig:quad-yplus-10-28}c-d). Comparing snapshots of inward interactions (Q3) in Fig. \ref{fig:vorticity}e-f, particles approach the wall with higher momentum in the positive streamwise direction where deposition occurs. Larger deceleration ejections away from the wall were also observed at these conditions (Fig. \ref{fig:vorticity}c-d). Ibrahim and Dunn \cite{Ibrahim2006} proposed that the detachment fraction of particles significantly larger ($d_p = 70 \mu$m) than studied in this work, may be reduced by a decline in ejection events in the near-wall region. While the combination of ejection-sweep events promote the transport of particles between bulk and wall regions, this suggests that ejections may also have a role in mitigating the deposition of coarse particles in the clean surface downstream. Furthermore, an increase in positive streamwise fluctuations leads to increased outward interactions in Fig. \ref{fig:vorticity}b that may also support mitigation.       

\subsection{Adverse effects of filtering}
\noindent Equipment filtering is traditionally used for removing coarse particle pollutants from the air before entry into the forced-cooling system (Fig. \ref{fig:9916-dust}a). Although this can reduce the concentration of coarse particles (PM$_{10}$), the additional flow resistance shifts the fan operating point in the direction of lower flow rates leading to reduced cooling efficiency. The effect of filter installation on the thermal resistance of four representative components located across the telecommunications base station equipment is listed in Table \ref{table:Rth}. Significant reductions of up to 30\% in thermal performance occur using a filter with an efficiency $\eta = 35-50\%$.

\begin{table}[b]
\centering
\caption{Changes to the thermal resistance of four sample components from PM$_{10}$ particle filtering}\label{table:Rth}
\begin{tabular}{l l l l l l l}
\hline
{} & $\eta (\%)$ & $\omega$ rad/s & $\Delta{R_1}/R_1$ & $\Delta{R_2}/R_2$ & $\Delta{R_3}/R_3$ & $\Delta{R_4}/R_4$ \\
\hline
MERV 5 & 20-35 & 419 & -0.121 & -0.098 & -0.100 & -0.240\\
MERV 5 & 20-35 & 497 & -0.096 & -0.083 & -0.059 & -0.210\\
MERV 6 & 35-50 & 419 & -0.195 & -0.184 & -0.143 & -0.302\\
MERV 6 & 35-50 & 497 & -0.151 & -0.156 & -0.040 & -0.254\\
\hline
\end{tabular}
\end{table}

Measurement of the exit flow revealed a change in the impingement angle up to $\approx 4^\circ$ due to increased levels of filtering. This trend, shown in Table \ref{table:dep}, also results from a shift in the fan operating point. Grimes \textit{et al.} \cite{Grimes2000} showed that shifting the fan operating point closer to stall conditions increases the spread of the conical jet at the outlet. Using a MERV 5 filter, the thermal performance of three out of four components is recovered to within 10\% of the baseline case (without a filter) by increasing the fan rotational speed from 419 rad/s to 497 rad/s. For $d_p = 2.5\mu$m, this increases $\tau^+$ by 10\% near the exit flow region which is sensitive to deposition. This change in both $\theta$ and $\tau^+$ shows that coarse particle filtering can indirectly enhance deposition velocities of fine particle pollutants by altering the time-averaged exit flow field.

\begin{table}[t]
\centering
\caption{The effect of PM$_{10}$ particle filtering on fine particle deposition mechanisms}\label{table:dep}
\begin{tabular}{l l l l l}
\hline
{} & $\eta (\%)$ & $\omega$ rad/s & $\theta (^\circ)$ & $\tau^+$\\
\hline
no filter & 0 & 419 & -18.8 & 0.379 \\
MERV 5 & 20-35 & 497 & -20.3 & 0.417 \\
MERV 6 & 35-50 & 497 & -22.6 & 0.399 \\
\hline
\end{tabular}
\end{table}

\begin{figure}[hb!]
\centering\includegraphics[width=0.8\linewidth]{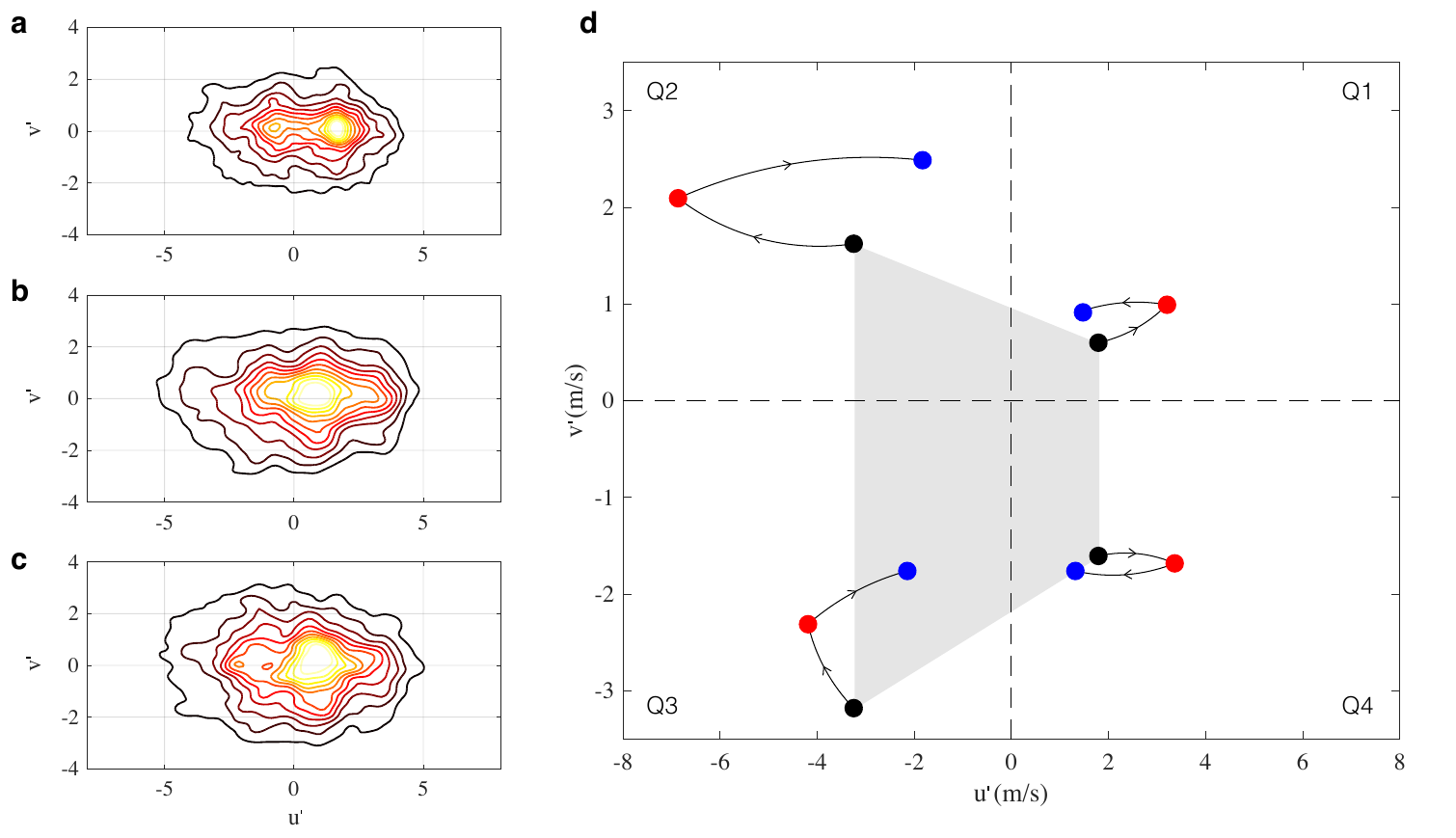}
\caption{Joint probability density functions for the cases a) without filtering, b) with $\eta  = 20-35\%$ and c) with $\eta  = 35-50\%$. The same location and wall distance of $y = 1.1$ mm was chosen across all cases.  Rotational speeds are a) 419 rad/s and b-c) 497 rad/s. d) Location of peaks in $u'v'P(u',v')$ for quadrants Q1-Q4. Symbols: $\circ$ (black) no filter; $\circ$ (red) $\eta  = 20-35\%$; $\circ$ (blue) $\eta  = 35-50\%$.}\label{fig:filter-JPDFs}
\end{figure}

Turbulent fluctuations ($u',v'$) are presented in Fig. \ref{fig:filter-JPDFs}a-c using joint probability density functions at the same near wall location for each case in Table \ref{table:dep}. A combination of filtering and an increase in rotational speed (to recover thermal performance) results in a wider spread in the fluctuations, most notably in the positive and negative streamwise directions. The overall Reynolds stress ($\overline{u'v'}$) increases by 40\% when using a MERV 5 filter. As before, the ejection-sweep events contribute most to this turbulence term. This increase in turbulent stress, combined with an increase in shear velocity and $\tau^+$, shows that adding coarse filtering increases the rate of deposition compared to the baseline case without a filter. At the same operating speed of 497 rad/s, the use of a MERV 6 filter reduces the fan flow rate even further. This reduction offsets an increase in impingement angle, leading to a reduction in the shear velocity and $\tau^+$ compared to the lower efficiency MERV 5 filter (Table \ref{table:dep}). This remains above the baseline values, however, and enhanced deposition rates would also be expected with this higher $\eta$ filter. Indeed, this scenario illustrates two negative impacts: 1) 15-25\% higher component thermal resistances and 2) increased deposition rates for fine particle pollutants and unfiltered coarse particles. 

The pairs of absolute velocity fluctuations that contribute most to the turbulent stress (quadrant peaks in $u'v'P(u'v')$) are shown in Fig. \ref{fig:filter-JPDFs}d for all three cases. The arrows indicate a move toward higher rotational speed and higher efficiency filtering. Sweep events (Q4) that promote particle transport to the surface have similar $-v'$ fluctuations. The low efficiency filter ($\eta = 20-35\%$) with increased rotational speed (497 rad/s) increases shear velocity and stretches all turbulent events in the positive and negative streamwise direction. This is the reason the highest turbulent stresses ($\overline{u'v'}$) occur for this case. Increasing the filtering efficiency ($\eta = 35-50\%$) and maintaining rotational speed (497 rad/s), reduces the fan flow rate, shear velocity and compresses the most stress-contributing turbulent events in the streamwise directions below even the low speed (419 rad/s) case without a filter. While these fluctuating pairs are in close proximity to the case without filtering, the joint probability density function in Fig. \ref{fig:filter-JPDFs}c shows that extreme fluctuations of similar magnitude to that found when using the lower efficiency filter also occur. These high-strength, less frequent turbulent eddy motions may also contribute to particle deposition over extended times. These \textit{in situ} observations on forced-cooled equipment highlight the sensitive dependency on the filter efficiency, fan rotational speed, performance, and the coupled relationship between these parameters on the particle flow field and deposition process.             

\section{Conclusions}
\label{Sec:Conc}
\noindent The deposition of coarse and fine particle pollutants from outdoor and poorly controlled indoor environments has become a major challenge in applications from equipment cooling to hygiene control. This paper experimentally investigated the advective-diffusive particle transport mechanisms that lead to preferential particle deposition downstream of turbulent axial fan flows. Field observations on forced-air cooled telecommunications equipment over long timescales ($\sim 1$ year) were investigated by performing particle image velocimetry \textit{in situ} using tracer particulates representing PM$_{2.5}$. The primary time-averaged mechanisms for the deposition pattern formation were found to be the oblique impingement created by the exit flow, together with localised changes to the shear velocity and particle relaxation time which ultimately lead to changes in deposition velocity. Using a quadrant analysis, the contributions of velocity fluctuations to the turbulent stresses were decomposed, indicating that sweeping events and the overall Reynolds stress have an important role in deposition through eddy diffusion and impaction. The flow fields of each turbulent event was visualised with instantaneous vorticity and velocity fields, isolated for pairs of velocity fluctuations that contribute most to turbulent transport in the near-wall region. From these observations, the effect of coarse particle filtering on fine particle transport was assessed. This revealed the strongly coupled effects of filtering on air flow distribution and particle deposition. Filtering generally promoted the mechanisms driving particle deposition, highlighting that filter selection in forced-air cooling systems requires careful considerations for the flow field, targeted filtering efficiencies, and particle composition of the environment. 


\end{document}